\documentclass[acmsmall,screen,nonacm]{acmart}
\usepackage{graphicx}
\usepackage{booktabs}
\usepackage{tabularx}
\usepackage{array}
\usepackage{enumitem}
\usepackage[dvipsnames]{xcolor}
\usepackage{framed}
\usepackage{changepage}
\usepackage{multirow}

\newcommand{\interviewquote}[2]{
\def\FrameCommand{
    \hspace{0pt} 
    {\color{RoyalBlue}  \vrule width 2pt}%
    {\color{white} \vrule width 3pt}%
  }%
  \MakeFramed{\advance\hsize-\width\FrameRestore}%
  \noindent
  \begin{adjustwidth}{}{1pt}
  {\footnotesize``#1'' - {#2}}\vspace{0.5pt}
  \end{adjustwidth}
  \endMakeFramed%
}

\definecolor{rulegrey}{HTML}{D9D9D9}

\newcolumntype{L}[1]{>{\raggedright\arraybackslash}p{#1}}
\newcolumntype{Y}{>{\raggedright\arraybackslash}X}
\newcolumntype{V}{!{\color{rulegrey}\vrule width 0.5pt}}

\usepackage{colortbl}
\usepackage{makecell}

\newcolumntype{S}{>{\raggedleft\arraybackslash}p{20pt}}

\newcolumntype{P}{>{\centering\arraybackslash}p{1.45em}}
\definecolor{dotgray}{gray}{0.80}
\definecolor{bandgray}{gray}{0.92}

\newcommand{\yes}{\textcolor{black}{$\bullet$}}
\newcommand{\no}{\textcolor{black!15}{$\bullet$}}
\usepackage[most]{tcolorbox}

\usepackage{tikz}

\usetikzlibrary{tikzmark,calc}

\definecolor{bandgray}{gray}{0.94}
\definecolor{rulegrey}{HTML}{D9D9D9}

\newlength{\spcol}    
\newlength{\panelcode}

\newcolumntype{S}{>{\centering\arraybackslash}p{\spcol}}

\definecolor{rqblue}{HTML}{D6E9F8}

\newtcolorbox{rqbox}[1]{
    enhanced,
    colback=rqblue!90,          
    frame hidden,
    arc=3pt,
    boxsep=0pt,                
    left=5pt, right=5pt, top=4pt, bottom=4pt,
    before skip=6pt, after skip=6pt,
    fontupper=\footnotesize,
}

\AtBeginDocument{%
  }

\begin{document}

\title[Permission Decisions in Agentic AI Assistants]{Let the Agent Do It? How Software Practitioners Understand and Make Permission Decisions in Agentic AI Assistants}

\author{Larissa Salerno}
\affiliation{%
  \institution{Chalmers University of Technology and University of Gothenburg}
  \city{Gothenburg}
  \country{Sweden}
}
\email{larissas@chalmers.se}

\author{Haoyu Gao}
\affiliation{%
  \institution{The University of Melbourne}
  \city{Melbourne}
  \country{Australia}
}
\email{haoyug1@student.unimelb.edu.au}

\author{Gregory Gay}
\affiliation{%
  \institution{Chalmers University of Technology and University of Gothenburg}
  \city{Gothenburg}
  \country{Sweden}
}
\email{ggay@chalmers.se}

\author{Alexander Serebrenik}
\affiliation{%
  \institution{Eindhoven University of Technology}
  \city{Eindhoven}
  \country{The Netherlands}
}
\email{a.serebrenik@tue.nl}

\author{Philipp Leitner}
\affiliation{%
  \institution{Chalmers University of Technology and University of Gothenburg}
  \city{Gothenburg}
  \country{Sweden}
}
\email{philipp.leitner@chalmers.se}

\renewcommand{\shortauthors}{Salerno et al.}

\begin{abstract}
Agentic AI assistants increasingly act on developers' behalf by modifying files, executing commands, and accessing external resources. These actions often require permission, yet little is known about how practitioners make permission decisions while still benefiting from agent autonomy. To address this gap, we conducted a sequential mixed methods study, interviewing 18 practitioners who use AI agents and then surveying 115 practitioners based on the interview findings. We find that practitioners often understand agent behaviour through what they can directly observe and review, while decisions, data use, and other activity behind the scenes remain less clear. This uncertainty also shapes permission decisions, which depend on the scope and risk of an action, whether it fits the task, familiarity with the agent, and the environment in which it operates. Practitioners respond by adjusting how closely they oversee agents, from setting limits in advance to monitoring execution and reviewing work afterwards. How much scrutiny they apply depends on factors such as trust, task importance, time pressure, and the consequences of an action. Our findings suggest that permission systems should make consequential actions easier to review, distinguish what an agent is allowed to do from what the user intended, make reversibility clearer, avoid treating repeated approvals as stable preferences, and distinguish rejecting a single action from rejecting an entire approach.
\end{abstract}



\maketitle
\section{Introduction}
\label{sec:introduction}

AI agents are being applied across software development workflows, including requirements analysis~\cite{oriol2025multi}, code generation~\cite{yang2024swe}, code review~\cite{lin2025codereviewqa}, and test generation~\cite{altmayer2025coverup}. This introduces a development paradigm where software practitioners communicate their goals through natural language and delegate the selection and execution of concrete actions to agents. Such delegation creates a need to manage permissions for actions chosen by the agent as a task unfolds. \textit{Runtime permission requests} ask software practitioners whether an agent may perform operations such as executing commands, modifying files, or accessing external resources, providing opportunities to control how delegated work proceeds~\cite{michael2026agents}.

Prior work has begun to explore permission mechanisms for AI agents and the challenges these tools introduce, from limited transparency in commercial approval workflows to the difficulty of anticipating what an agent will do next~\cite{michael2026agents}. A separate body of research on human factors in security and automation shows that presenting a prompt does not guarantee that people understand or act on it, and describes broader forms through which people oversee automated systems~\cite{felt2012android,parasuraman1997humans,dhanorkar2026human}. Indeed, the mechanisms that agents deploy in practice all route decisions through the user, which has led to calls to treat agent security as a question of agent-human interaction rather than of enforcement alone~\cite{wang2026reframing}. However, little is known about how software practitioners themselves interpret and respond to individual permission requests from coding agents as a task unfolds in practice, including what they understand about a requested action, how they decide whether to grant it, and how they oversee an agent's behaviour once permission has been given.

We address this gap through a mixed methods study of how software practitioners understand agentic AI assistants and make and reflect on permission decisions from these assistants in their daily work. We conducted a two-phase sequential study consisting of an interview study followed by a survey. In the first phase, we interviewed software practitioners with experience using agentic AI assistants (e.g., Claude Code, Codex, Gemini CLI) about their use and understanding of these tools, their permission decisions, and their oversight practices. We qualitatively analysed the interview data to identify recurring behaviours, perceptions, decision factors, permission boundaries, and oversight practices. In the second phase, we developed a survey informed by the interview findings to validate the patterns identified in the interviews with a broader sample of software practitioners. Combined, our study addresses the following research questions:

\begin{rqbox}{}
\noindent\textbf{RQ1. How do software practitioners understand the actions of agentic AI assistants?} 
\end{rqbox}

\noindent We focus on the extent to which practitioners understand how these agents operate and act on their machines, with a particular focus on permission-related actions and requests.  We find that practitioners build their understanding mainly from what they can observe and review, such as outputs, changes, logs, and reasoning traces, while still facing gaps in visibility and uncertainty about what agents do behind the scenes.

\begin{rqbox}{}
\noindent\textbf{RQ2. How do software practitioners make decisions about permission requests from agentic AI assistants?}
\end{rqbox}

\noindent We look into how practitioners respond when they receive permission requests, the types of permissions they are willing to grant, and the factors behind these decisions. We find that permission decisions depend strongly on context, including the risk and scope of the action, whether it fits the task, and the environment in which the agent is running. Participants are generally more comfortable with bounded and low-risk access, while sensitive data, privileged access, destructive actions, and actions outside the intended scope form stronger boundaries.

\begin{rqbox}{}
\noindent\textbf{RQ3. How do practitioners oversee and reflect on the actions and permissions of agentic AI assistants?}
\end{rqbox}

\noindent We investigate how practitioners monitor agent actions and permissions, respond to unexpected behaviour, and assess the transparency of these tools. We find that practitioners oversee agents in different ways, including setting constraints in advance, following execution as it happens, and reviewing the result afterwards, with the level of oversight changing with factors such as risk, task importance, trust, and time pressure.

These findings suggest that permission systems should do more than simply ask users to approve individual actions. Tools should make important actions and resource access easier to review, account for the fact that repeated approval may reflect fatigue rather than preference, and make the consequences and reversibility of actions clearer. They should also distinguish between what an agent is technically allowed to do and what the user actually intended it to do, while giving users ways to distinguish between denying a single action and ruling out an entire approach.
\section{Related Work}\label{sec:related-work}

\subsection{Agentic AI in Software Development}

Coding agents enable practitioners to automate implementation tasks. Systems such as SWE-agent navigate repositories, edit files, and execute tests in response to development tasks~\cite{yang2024swe}. In a field study at Microsoft, Murphy-Hill et al. found that adopting Claude Code and GitHub Copilot CLI was associated with sustained increases in developers' merged pull requests~\cite{murphy2026adoption}. These capabilities support different ways of organising development. In its original formulation, ``\textit{vibe coding}'' involves delegating implementation through natural-language prompts without directly reviewing or editing the generated code~\cite{sarkar2025vibe}. In practice, they found that vibe coding involved iterative cycles of prompting, inspecting generated code, testing applications, and manually editing code. More recently, ``\textit{spec-driven development}'' has emerged as a more structured workflow for guiding agent implementation, in which developers write or refine structured natural-language specifications that direct agents' work~\cite{agarwal2026specmine}.

The use of AI in software development introduces challenges regarding productivity, understanding, and software quality. Becker et al. found that early-2025 AI tools slowed experienced developers working in familiar repositories despite perceived productivity gains~\cite{becker2025measuring}. Balepur et al. observed improved initial task completion but weaker code comprehension among students using coding agents~\cite{balepur2026paired}. Baltes et al. examined developers' discussions of AI ``slop'', identifying concerns about review burden, code quality, and the wider costs of low-quality AI-generated contributions~\cite{baltes2026endless}. Gao et al. found that AI-co-authored pull requests were merged faster and received fewer recorded comments and reviews than human-only contributions~\cite{gao2026autopilot}. Choudhuri et al. further found that developers wanted AI support with clearly defined limits on its authority, while retaining human judgement and control over consequential decisions~\cite{choudhuri2026copilot}. These findings motivate closer examination of how developers understand and oversee AI-assisted work.

\subsection{Permission Mechanisms in Agentic AI}

\textit{Permission mechanisms} govern which actions agents may perform on a user's behalf. Michael and Roesner surveyed 21 permission system proposals and compared five commercial agents, finding that commercial systems often require choosing between burdensome manual approvals and automated permission decisions made by LLMs, with limited transparency and user control~\cite{michael2026agents}. Kravchenko et al. proposed an information-flow control framework that checks whether tool calls comply with security policies before execution and verifies their outputs before admitting them into the agent's context. It can process untrusted data in disposable branches and return validated structured results while discarding the branch transcript, allowing the main agent to continue without incorporating the raw untrusted content~\cite{kravchenko2026agentic}. Shukla et al. proposed Hedwig, which learns behavioural guidelines from developers' decisions and feedback to adjust when coding agents seek further input, accommodating changes in oversight preferences across tasks and over time~\cite{shukla2026hedwig}. Wu et al. conducted a user study to identify contextual factors shaping data-access permission decisions, then combined contextual information and prior decisions to predict future choices~\cite{wu2026towards}. Their approach automates permission handling while accounting for differences in users' preferences. 

Other work examines how the timing and granularity of these decisions affects their effectiveness. Zhou et al. modelled confirmation as a scheduling problem to place checkpoints inside a task, and found that 48 participants preferred intermediate confirmation over the confirm-at-end approach of existing systems~\cite{zhou2026confirmation}. Yan compared per-action approval, automated model review, and user-authored consequence policies across 113 participants supervising a simulated day of agent actions. Counter-intuitively, the user-authored policies blocked less overreach than either per-action alternative, and most overreach actions were explicitly approved by a human at runtime~\cite{yan2026policies}. Surveying 59 papers, 21 production systems, and 26 security plug-ins, Wang et al. argued that agent security is primarily an agent-human interaction problem, as the mechanisms industry deploys leave users caught between approval fatigue and uncontrolled agent autonomy~\cite{wang2026reframing}.

Developers also guide agents through skills and repository context. Gao et al. found that skills are commonly copied, with behavioural constraints remaining largely stable as operational instructions are adapted~\cite{gao2026registry}. Treude et al.'s preliminary investigation illustrates how developers encode privacy, fairness, and accountability requirements in repository context files~\cite{treude2026operationalizing}. Such artefacts express expectations before execution, while runtime permission requests ask developers to judge specific actions. These studies explain how developers express expectations for agents, but offer limited insight into how they interpret and respond to permission requests during execution.

\subsection{Human Permission Decisions and Oversight of Automation}

Research on usable security highlights the relevance of user attention, context, and interface design to how users respond to permission requests and security warnings~\cite{acquisti2017nudges}. Studies of Android permissions report limited attention and comprehension~\cite{felt2012android}, alongside context-dependent preferences for resource access~\cite{wijesekera2015android}. Research on browser warnings and security dialogues shows that presentation affects user responses~\cite{akhawe2013alice,bravo2013your}, while repeated exposure can reduce attention through habituation~\cite{vance2018tuning}. Research on automation further distinguishes appropriate reliance from misuse and disuse~\cite{parasuraman1997humans,lee2004trust}. Together, these studies suggest that providing a permission prompt does not ensure that users understand the request or assess its consequences.

Dhanorkar et al. interviewed developers and identified four forms of agent oversight: a priori control, co-planning, real-time monitoring, and post-hoc review~\cite{dhanorkar2026human}. Consistent with post-hoc review as a form of oversight, eye-tracking evidence shows that although developers do not review LLM-labelled code more thoroughly overall, they fixate longer on it and adapt their review strategy, revealing a gap between reviewers' stated intentions and their actual behaviour~\cite{Khojah2026samescrutiny}. While the Android permission studies above examined access to protected resources, coding agent requests can involve commands selected by the agent as a task unfolds~\cite{michael2026agents}. Developers must assess both the consequences of an operation and whether it fits the task they delegated. This creates a tension between delegating implementation and retaining enough understanding to authorise its individual steps. How software practitioners make these assessments when responding to individual permission requests remains insufficiently understood.

Altogether, this body of work shows what agentic coding tools can do and the challenges they introduce, early attempts to design permission mechanisms for agents, and general principles of human oversight of automation. However, no study has looked into how developers understand and decide on individual permission requests from coding agents as they arise during development. This paper addresses this foundational gap by studying how software practitioners understand and act on permission requests spanning file modifications, command execution, and data access and sharing, and how they oversee agent actions during delegated development.
\section{Methodology}
\label{sec:methodology}

\begin{figure}[t]
    \centering
    \includegraphics[width=0.95\textwidth]{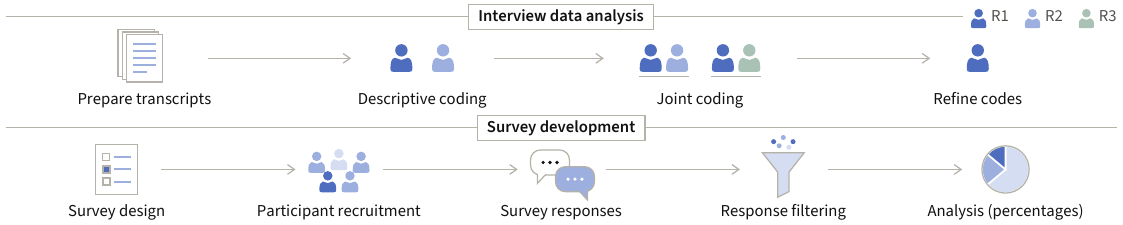}
    \Description{Study design process. The top row shows the interview analysis in four steps: preparing transcripts, descriptive coding by two researchers, joint coding by two pairs of researchers, and refining codes. The bottom row shows the survey in five steps: survey design, participant recruitment, survey responses, response filtering, and analysis.}
    \caption{Overview of the interview analysis and survey study, from transcript preparation and coding to survey design, participant recruitment, response filtering, and analysis. R1--R3 refer to the three researchers involved in the interview analysis.}
    \label{fig:dataanalysis}
\end{figure}

Figure~\ref{fig:dataanalysis} provides a visual overview of our research methods. We conducted this research as a sequential mixed methods study~\cite{storey2025guiding} with two steps: (1) a qualitative interview study, and (2) a confirmatory web-based survey intended to quantify the findings from the first step. This method follows established empirical research practice in software engineering, and has been used extensively in earlier work~\cite{cito2015making,Erlenhov2020fse,bacchelli2013expectations,wan2021machine,bi2022accessibility}.
Eligible participants for both phases were software practitioners with experience using AI agents, and participation was voluntary and uncompensated throughout. Under the first author's institutional procedures and legal requirements, neither phase of this study required formal ethical review.

\subsection{Interviews}

We recruited software practitioners through a combination of convenience and purposive sampling \cite{baltes2022sampling}. Our professional network provided an accessible pool, while we purposively sought variation in participants' roles, experience, AI-agent use, and countries of work. Country variation was intended to capture potentially different regulatory contexts, although the study was not designed for country-level comparisons. Prospective participants completed a background questionnaire, and eligibility was assessed manually. We recruited 18 respondents meeting these criteria. Rather than a predefined sample size, we used saturation as a stopping criterion: few new codes emerged after P12, and P17 and P18 introduced no new codes, suggesting that code saturation had been reached by the final interviews. 

\newcommand{\nodata}{\multicolumn{1}{c}{--}}

\begin{table*}[t]
\centering
\caption{Participant background characteristics for the interview ($N = 18$) and survey ($N = 115$) samples. Filled dots indicate interview participant responses, $n$ gives interview counts, and SP\% indicates survey percentages. A dash indicates that the item was not asked in the survey. Participants could select more than one AI assistant tool. All other questions were limited to a single selection.}
\label{tab:demographics}
\setlength{\tabcolsep}{3pt}
\renewcommand{\arraystretch}{1.05}
\tiny
\begin{tabular*}{\textwidth}{@{\extracolsep{\fill}}lcccccccccccccccccc|r|r@{}}
\toprule
\textbf{Question/Options} & \textbf{P01} & \textbf{P02} & \textbf{P03} & \textbf{P04} & \textbf{P05} & \textbf{P06} & \textbf{P07} & \textbf{P08} & \textbf{P09} & \textbf{P10} & \textbf{P11} & \textbf{P12} & \textbf{P13} & \textbf{P14} & \textbf{P15} & \textbf{P16} & \textbf{P17} & \textbf{P18} & \textbf{$n$} & \textbf{SP\%} \\
\midrule
\rowcolor{black!8} \multicolumn{21}{@{}l}{\textit{Which country do you work in?}} \\
Sweden & \yes & \yes & \no & \yes & \yes & \yes & \yes & \no & \yes & \no & \yes & \no & \no & \no & \no & \yes & \no & \no & 9 & 27.8\% \\
Brazil & \no & \no & \no & \no & \no & \no & \no & \yes & \no & \yes & \no & \no & \no & \no & \no & \no & \no & \yes & 3 & 25.2\% \\
China & \no & \no & \no & \no & \no & \no & \no & \no & \no & \no & \no & \no & \yes & \yes & \yes & \no & \no & \no & 3 & 0.0\% \\
Netherlands & \no & \no & \no & \no & \no & \no & \no & \no & \no & \no & \no & \yes & \no & \no & \no & \no & \yes & \no & 2 & 3.5\% \\
Australia & \no & \no & \yes & \no & \no & \no & \no & \no & \no & \no & \no & \no & \no & \no & \no & \no & \no & \no & 1 & 2.6\% \\
Other & \no & \no & \no & \no & \no & \no & \no & \no & \no & \no & \no & \no & \no & \no & \no & \no & \no & \no & 0 & 40.9\% \\
\addlinespace[1pt]
\rowcolor{black!8} \multicolumn{21}{@{}l}{\textit{How many years of professional experience do you have in IT?}} \\
0--2 years & \yes & \no & \no & \no & \no & \no & \no & \no & \no & \no & \no & \no & \no & \yes & \no & \no & \no & \no & 2 & 9.6\% \\
3--5 years & \no & \yes & \no & \yes & \no & \no & \no & \no & \no & \no & \no & \no & \yes & \no & \no & \no & \no & \no & 3 & 20.0\% \\
6--10 years & \no & \no & \yes & \no & \yes & \yes & \no & \yes & \yes & \yes & \yes & \yes & \no & \no & \yes & \no & \yes & \yes & 11 & 27.0\% \\
11--15 years & \no & \no & \no & \no & \no & \no & \yes & \no & \no & \no & \no & \no & \no & \no & \no & \yes & \no & \no & 2 & 22.6\% \\
16+ years & \no & \no & \no & \no & \no & \no & \no & \no & \no & \no & \no & \no & \no & \no & \no & \no & \no & \no & 0 & 20.9\% \\
\addlinespace[1pt]
\rowcolor{black!8} \multicolumn{21}{@{}l}{\textit{What is your seniority level?}} \\
Junior & \yes & \yes & \no & \no & \no & \no & \no & \no & \no & \no & \no & \no & \no & \yes & \no & \no & \no & \no & 3 & \nodata \\
Mid-level & \no & \no & \yes & \yes & \yes & \no & \no & \no & \no & \no & \no & \yes & \yes & \no & \no & \no & \no & \yes & 6 & \nodata \\
Senior & \no & \no & \no & \no & \no & \yes & \yes & \yes & \yes & \yes & \yes & \no & \no & \no & \yes & \yes & \yes & \no & 9 & \nodata \\
\addlinespace[1pt]
\rowcolor{black!8} \multicolumn{21}{@{}l}{\textit{How large is the company you work for?}} \\
Startup (1--50) & \no & \no & \no & \no & \no & \no & \no & \no & \no & \no & \no & \yes & \no & \no & \no & \no & \no & \no & 1 & \nodata \\
SME (51--500) & \no & \yes & \no & \no & \no & \yes & \no & \no & \yes & \no & \no & \no & \no & \no & \no & \no & \no & \yes & 4 & \nodata \\
Large enterprise (500+) & \yes & \no & \yes & \yes & \yes & \no & \yes & \yes & \no & \yes & \yes & \no & \yes & \yes & \yes & \yes & \yes & \no & 13 & \nodata \\
\addlinespace[1pt]
\rowcolor{black!8} \multicolumn{21}{@{}l}{\textit{How long have you been using AI assistant tools?}} \\
Less than 6 months & \no & \no & \no & \yes & \no & \no & \no & \yes & \no & \no & \no & \no & \no & \no & \no & \no & \no & \yes & 3 & 5.2\% \\
6--12 months & \yes & \no & \no & \no & \yes & \no & \no & \no & \no & \no & \yes & \yes & \yes & \yes & \no & \no & \yes & \no & 7 & 33.9\% \\
1--2 years & \no & \yes & \yes & \no & \no & \yes & \yes & \no & \no & \yes & \no & \no & \no & \no & \yes & \yes & \no & \no & 7 & 41.7\% \\
More than 2 years & \no & \no & \no & \no & \no & \no & \no & \no & \yes & \no & \no & \no & \no & \no & \no & \no & \no & \no & 1 & 19.1\% \\
\addlinespace[1pt]
\rowcolor{black!8} \multicolumn{21}{@{}l}{\textit{How often do you use AI assistant tools?}} \\
Daily & \no & \yes & \yes & \no & \yes & \yes & \yes & \yes & \yes & \yes & \no & \yes & \yes & \yes & \yes & \no & \yes & \yes & 14 & \nodata \\
Weekly & \yes & \no & \no & \yes & \no & \no & \no & \no & \no & \no & \no & \no & \no & \no & \no & \yes & \no & \no & 3 & \nodata \\
Occasionally & \no & \no & \no & \no & \no & \no & \no & \no & \no & \no & \yes & \no & \no & \no & \no & \no & \no & \no & 1 & \nodata \\
\addlinespace[1pt]
\rowcolor{black!8} \multicolumn{21}{@{}l}{\textit{Which AI assistant tools do you use? (tick all that apply)}} \\
Claude Code & \yes & \no & \yes & \yes & \yes & \no & \no & \yes & \yes & \yes & \no & \yes & \no & \yes & \yes & \no & \yes & \yes & 12 & \nodata \\
GitHub Copilot & \yes & \yes & \no & \no & \no & \no & \yes & \no & \yes & \no & \yes & \no & \no & \no & \no & \yes & \yes & \no & 7 & \nodata \\
Codex & \yes & \no & \no & \no & \no & \no & \no & \no & \no & \no & \no & \yes & \yes & \yes & \no & \no & \no & \no & 4 & \nodata \\
Gemini CLI & \no & \no & \yes & \no & \no & \no & \no & \no & \yes & \no & \no & \no & \no & \no & \no & \yes & \no & \no & 3 & \nodata \\
Cursor & \no & \yes & \no & \no & \no & \yes & \no & \no & \no & \no & \no & \no & \no & \no & \no & \no & \no & \no & 2 & \nodata \\
Other tools & \no & \no & \no & \no & \no & \no & \no & \no & \no & \yes & \no & \no & \no & \no & \no & \no & \yes & \yes & 3 & \nodata \\
\bottomrule
\end{tabular*}
\end{table*}

Table \ref{tab:demographics} summarises the participants' professional and AI-tool-use backgrounds.

We conducted semi-structured interviews between June and August 2026. Seventeen interviews were conducted online using Microsoft Teams, while one was conducted in-person and recorded using Teams.

The interview guide was developed after two trial interviews with one co-author and one industry contact, and included questions about what participants use AI agents for, how well they understood the actions taken by agents on their computers, and how they responded to permission requests. No substantive changes were made to the guide during data collection. The interview guide is available in our online appendix (see Section \ref{sec:dataavailability}).

Interviews typically lasted approximately 30--45 minutes. One participant's interview was substantially longer and was continued in a second session on a separate day. Interviews were conducted in English, Portuguese, or Chinese by an author fluent in this language. Non-English interview transcripts were machine-translated into English and subsequently checked by an author fluent in the source language as well as English. Participants provided written informed consent before each interview. We did not report participants' names or company identities and do not make the interview transcripts publicly available to protect participant confidentiality, following recommendations for ethical interview research in software engineering \cite{strandberg2019ethical}. The consent form is available in our online appendix (see Section \ref{sec:dataavailability}).

We analysed the interview data using a multi-stage coding and categorisation process, drawing on open coding and constant comparison procedures described in Socio-Technical Grounded Theory \cite{hoda2022stgt}. We identified excerpts relevant to the interview topics and summarised them using descriptive codes, which were refined into shorter focused codes covering participants' understanding of agent behaviour, responses to permission requests, and oversight practices. To align their coding approach, two authors first coded two transcripts jointly. One author then coded ten transcripts independently and the other five, while the first author and a third, senior author jointly coded one. To check the consistency of the coding, the third author also independently re-coded two transcripts, and the authors compared and discussed differences. The first author then reviewed the focused codes across all participants, iteratively merging, renaming, or consolidating codes that expressed similar ideas to reduce overlap and produce a final set of codes. Related final codes were grouped into higher-level categories and dimensions, developed from the data rather than defined in advance, which we used to structure the results. Our full codebook is available online (Section \ref{sec:dataavailability}).

\subsection{Survey}

Following the interview analysis, we developed a confirmatory survey to explore whether the patterns identified in the interviews were also recognised by a broader sample of software practitioners. The survey covered the same topics as the interviews, i.e., how participants understand AI agents, how they respond to permission requests, the factors that influence their permission decisions, and their perceptions of permissions and agent transparency. Response options were based on the final interview codes. For categorical questions, we kept all categories, and for permission boundaries, the specific permission or action types. For other topics, we selected recurring, clearly distinct findings that could be turned into structured survey items, while leaving out highly specific, overlapping, or difficult-to-operationalise findings to keep the survey manageable. Items included multiple-choice questions (both select one and select all that apply), statement selections, and several with an optional open-text ``Other'' response. One instructed-response attention check was included to help identify inattentive responses~\cite{meade2012identifying}. The survey was designed to take approximately 8--10 minutes to complete. We pre-tested the survey with three PhD students, whose feedback led us to revise wording, add the attention-check item, and remove one question on unexpected agent behaviour because the predefined response options felt too restrictive for that topic.

Participants were first presented with a page describing the purpose of the study, the expected survey length, the anonymous nature of participation, how their responses would be used, and how to contact the research team. They were informed that they could stop the survey at any time before submitting their responses, and that continuing with the survey indicated their consent to participate. We administered the survey through Typeform~\cite{typeform}. Similar to interview transcripts, we do not make individual-level survey responses publicly available and instead report findings in aggregate form. Our survey instrument is also contained in our online appendix (Section \ref{sec:dataavailability}).

We recruited participants using a combination of convenience, purposive, and snowball sampling, distributing the survey through the research team's industry contacts, relevant Reddit communities focused on software development and AI agents, LinkedIn, Bluesky, and Mastodon, and by asking interview participants to share the survey within their organisations or professional networks. We considered the online channels appropriate for reaching software practitioners, as previous work has identified active software development communities on Reddit and shown that software engineering practitioners engage with research on LinkedIn~\cite{aniche2018modern,wyrich2024beyond}. Eligibility was assessed from participants' self-reported job title or role and their reported length of AI-agent use. Responses from participants who did not meet these eligibility criteria, failed the attention check, or were incomplete were excluded from the analysis. We also included a screening question asking whether participants had noticed AI agents requesting permission before taking actions or accessing resources. Participants who answered ``No'' were redirected to the end of the survey, while those who answered ``Yes'' or ``I am not sure'' continued with the permission-focused questions. 

We, therefore, use two survey samples in the analysis: 115 participants for questions appearing before and including the screening question and 107 participants for the permission-focused questions. Column \emph{SP\%} in Table~\ref{tab:demographics} reports background characteristics of survey respondents. Because the survey collected fewer background items than the interviews, percentages are shown only for country, professional IT experience, and length of AI-agent use. For country, ``Other'' groups all countries not listed individually in the table, and all survey respondents reported a country. We calculated response frequencies for each survey question. Percentages are based on the number of participants eligible to answer each question. Responses entered through the optional open-text ``Other'' fields were reviewed manually. As relatively few were provided (21 responses across 7 questions), relevant responses were summarised in the text. The figures reporting survey results were generated with Python scripts written with the assistance of an AI tool (Claude); all plotted values are computed from the survey response counts and were verified by the authors against the survey analysis. The survey results were organised around the same three research questions as the interview findings, and we compared the survey responses with the interview findings to see whether similar behaviours and perceptions were reported by the broader sample.
\section{Results}
\label{sec:results}

We organise the results around the three research questions. We find that practitioners' understanding of agent behaviour is uneven and largely confined to what they can directly observe, such as outputs, changes, and reasoning traces (\textbf{RQ1}). Because this understanding is partial, permission decisions rely on heuristics such as scope, risk, familiarity, and task fit rather than a full assessment of what an agent is about to do (\textbf{RQ2}). These heuristics, together with the visibility gaps that limit understanding, in turn shape how practitioners oversee agents once permission has been granted, producing oversight strategies that shift with task importance and risk, task complexity, demonstrated reliability, and time pressure (\textbf{RQ3}). The labels in the figures in the Results section represent findings identified in the interviews, while the percentages show how many survey respondents selected the corresponding survey item. Only interview findings that were included in the survey are shown in these figures.

\subsection{Understanding of Agentic AI Assistant Behaviour (RQ1)}
\label{sec:RQ1}

\begin{figure}[t]
    \centering
    \includegraphics[width=1.0\textwidth]{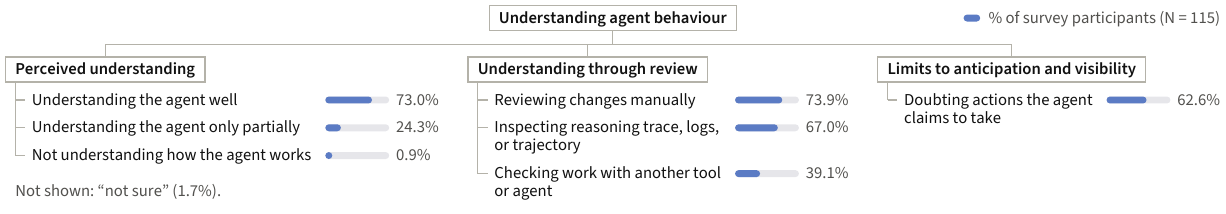}
    \Description{Tree diagram with categories showing the percentage of 115 survey participants who selected each statement about their perceived understanding of AI agents, how they review agent work, and doubts about actions agents claim to take.}
    \caption{Participants' perceived understanding of agent behaviour, how they review agent work, and the limits to anticipating agent actions (N = 115).}
    \label{fig:understanding}
\end{figure}

We examined participants' understanding of agent behaviour, which helps explain their approaches to permission and oversight. We organise our findings into four categories, and Figure \ref{fig:understanding} summarises the three categories directly covered by the survey. \textit{Perceived understanding} represents participants' self-assessment. \textit{Understanding through review} describes how participants improve or verify that understanding. \textit{Trust and sharing judgements} reflects how participants reason about information shared with agents and whether they trust agents or their operational environment. Finally, \textit{limits to anticipation and visibility} outlines situations where participants struggle to predict, observe, or maintain visibility of agent actions.

\smallskip\noindent\textbf{Perceived understanding.} We distinguish three levels of understanding: full, partial, and none. Almost half our interviewees described partial or no understanding. Conversely, survey respondents judged their understanding substantially more positive: 73.0\% believe they have full understanding, 24.3\% partial, and only 0.9\% report no understanding (1.7\% were unsure).\footnote{However, recall that the survey results reflect \emph{self-assessed} understanding, which is typically higher than real understanding of a topic \cite{9402043}.}

Confidence is often confined to what participants can directly observe, and what happens behind the scenes is less clear. Some interviewees feel they understand systems better when they can see how the underlying process works, but are less sure how tools like Claude Code decide what to do, when to request permission, or how data is used:

\interviewquote{\textit{Under the hood, no, not not really. I only understand the output of what it's giving me and only observable things}}{P05}

\noindent Technical knowledge does not necessarily make these behaviours clearer:

\interviewquote{\textit{I understand some of the basic principles of machine learning [...] but for example, what Visual Studio Code is doing under the hood [...] is something that I need to understand better}}{P16}

\noindent These findings confirm concerns raised in prior work: greater agent autonomy can make it harder for users to understand both the resulting code \cite{balepur2026paired} and the agent's behaviour \cite{chen2026code}.

\smallskip\noindent\textbf{Understanding through review.} Participants often try to understand what the agent was doing by examining its work (output) during or after a task. P03, for example, gains confidence by inspecting the agent's trajectory, and checking which tools and scripts it invokes. P14 similarly relies on seeing the changes the agent makes and becomes more cautious when they are not shown:

\interviewquote{\textit{If I find that it claims to have done something but doesn't show me the changes, that's when I step in}}{P14}

\noindent P17 similarly finds it easier to understand when they can follow their agent as it works:

\interviewquote{\textit{If [...] 
I only see the final result. I find it hard to understand what it did}} {P17} 
 
Others apply tools or other agents to double-check work. P14 asked a second agent to assess another agent's conclusions, while P15 described using another agent or tooling to check generated code for code smells. 

73.9\% of survey respondents manually review the agent's changes, 67.0\% check the reasoning trace, logs, or trajectory, and 39.1\% use another tool or agent. Textual responses mentioned skimming changes to catch cases where the agent had over-reached and running code in test environments. 

Participants figure out what the agent has done by checking its work afterwards, rather than already knowing how it works up front, similar to the real-time monitoring and post-hoc review described by Dhanorkar et al.~\cite{dhanorkar2026human}. Here, these practices served both as review and to build understanding.

\smallskip\noindent\textbf{Trust and sharing judgements.} Interviewees considered where their data might go and their comfort in sharing it. For some, this judgement comes before any question of permissions:

\interviewquote{\textit{Before asking something from Claude or anything else, I consider whether I'm okay sharing this data [...] If it's something sensitive, of course, I just don't use it for that}}{P05}

Survey respondents share similar feelings: 61.7\% would never grant an agent access to sensitive data, and 58.9\% said the same for email or other personal information (see Figure \ref{fig:permissionboundaries}). This is consistent with work showing that data sensitivity strongly shapes access decisions for agents \cite{wu2026towards}.

Once P05 considers a project safe to share, they are less concerned about what happens ``under the hood''. For other interviewees, how data is used still matters. P08 considers where their code is sent and whether it could be used for model training, while P11 feels more comfortable when models are hosted within the company's own infrastructure. Sharing decisions depend not only on data sensitivity, but also where it is processed and how it will be handled. Dhanorkar et al. similarly found that developers were concerned about data sent to third-party platforms~\cite{dhanorkar2026human}.

\smallskip\noindent\textbf{Limits to anticipation and visibility.} Reviewing agent actions does not mean participants can anticipate or verify actions. 62.6\% of survey respondents expressed doubt in actions agents claimed to be taking. One response said that, despite understanding the technical setup, they sometimes doubt whether agents do what they claim and would step in to guide them. P12 also described approving an action based on one expectation, only for the agent to do something different.

Visibility varies between tools. P04 feels they have more control over permissions and data flow with open-source frameworks than with Claude Code, while P11 feels there is little visibility into background activity, especially when agents work simultaneously: 

\interviewquote{\textit{What the sub-agents are actually doing and what instructions they are using, honestly, you can't really perceive that}}{P11} 

\noindent This challenge may become more pronounced as coding tools increasingly support subagents and background tasks that run in parallel \cite{anthropic2025autonomous}.

\begin{rqbox}{Understanding of Agent Behaviour (RQ1):}
\textit{\textbf{Takeaway:}} Software practitioners build understanding through what they can observe and review: changes, reasoning traces, logs, and outputs. Although a majority feel they understand agent output, they express doubts regarding visibility of individual decisions, trust, and access to sensitive or personal data.
\end{rqbox}

\subsection{Permission Decision-Making (RQ2)}
\label{sec:RQ2}

\begin{figure}[t]
    \centering
    \includegraphics[width=1.0\textwidth]{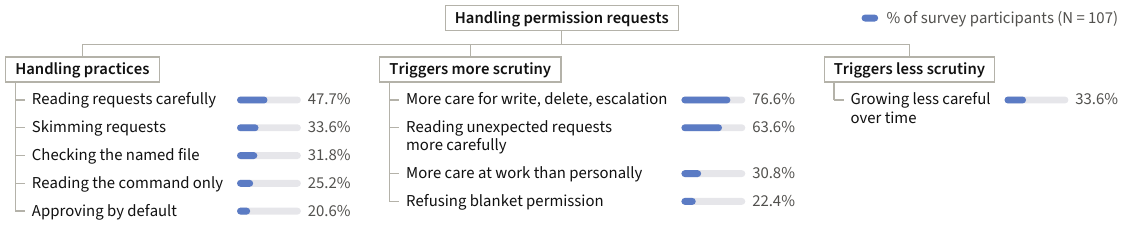}
    \Description{Tree diagram with categories showing the percentage of 107 survey participants who reported each way of handling permission requests and each situation that makes them more or less careful.}
    \caption{How participants handle permission requests, including their handling behaviours and the conditions that lead to more or less scrutiny.}
    \label{fig:handlingpermissions}
\end{figure}

Because participants' understanding of agent behaviour is often partial and limited to what they can observe (\textbf{RQ1}), permission requests become a key point at which they can actively shape what an agent is allowed to do. We now discuss how software practitioners react to agent permission requests.

\subsubsection{\textbf{Handling Permission Requests}}

We organise findings regarding review and approval practices (Figure \ref{fig:handlingpermissions}) into three groups: \textit{handling practices} reflects participants' reactions to permission requests, while \textit{triggers more scrutiny} and \textit{triggers less scrutiny} describe factors that make participants more or less careful.

\smallskip\noindent\textbf{Handling practices.} Permission requests are handled in various ways. 47.7\% of our survey respondents believe they read permission requests carefully. P12 describes how tiring this can become:

\interviewquote{\textit{[I] pay attention to what it's asking permission for, and then manually approve or not. This really is exhausting.}}{P12}

Reviewing a request does not necessarily mean reading it in full. P03, P05, and P14 skim requests, as do 33.6\% of survey respondents. Participants also often focus on particular aspects of a request: 31.8\% check the named file, while 25.2\% focus on the command or title. P05 reads more carefully when something is unfamiliar.

Some interviewees approved requests by default, an approach also reported by 20.6\% of survey respondents, while others prefer blanket permissions. However, some reported using blanket approvals while still keeping certain tools, commands, credentials, or sensitive code off-limits.
These approaches reflect a trade-off between control and efficiency, as reviewing each request takes time and attention. As stated by P17:

\interviewquote{\textit{Otherwise it kind of defeats the point of being faster, because every 10 seconds I have to click something}}{P17}

\smallskip\noindent\textbf{Triggers more scrutiny.} Participants increase their scrutiny when requests carry more risk, are unexpected, or happen in less familiar or controlled settings. P09 highlighted write requests:

\interviewquote{\textit{When it's trying to write, I pay much more, much closer attention, especially if it's something that I don't expect}}{P09}

P05 and P07 also noted that they apply more scrutiny to unexpected requests. In the survey, 76.6\% said they check write, delete, or privilege-escalation requests more carefully, while 63.6\% do so when a request is unexpected. Context matters, too: P02, P04, and P16 are more careful at work than on personal projects, while P12 is more cautious outside a controlled environment. 30.8\% of survey respondents also said they are more careful at work than on personal projects. In addition, 22.4\% of survey respondents refuse blanket permission and prefer to be asked each time. This preference for case-by-case handling is also visible when users are offered an alternative: given the chance to set standing rules in advance, participants in an earlier study chose ``\textit{ask}'' for most of them, returning the decisions to runtime \cite{yan2026policies}.

\smallskip\noindent\textbf{Triggers less scrutiny.} 
The main factor that triggers participants to be less careful is familiarity: in the survey, 33.6\% of respondents said they have become less careful about permission requests over time (and, given that this is self-reported, this number is likely an underestimation). P04 describes clicking through repeated prompts without looking at them closely, while P17 sometimes moves straight to bypassing permissions.

Having to deal with many requests can have a similar effect: at some point, there are simply too many to review individually.

\interviewquote{\textit{There's no way that I can actually go through all the permission requests}}{P03}

This is similar to habituation to repeated security warnings \cite{vance2018tuning} and approval fatigue reported for Claude Code, where users approve 93\% of permission prompts \cite{hughes2026automode}. Wang et al. describe a trade-off between cognitive burden and security guarantees that current mechanisms leave unresolved~\cite{wang2026reframing}.

P03, P06, and P13 also pay less attention when working in a controlled environment. They rely on the sandbox around the agent, giving agents more autonomy within the defined workspace while requiring additional approval beyond it~\cite{dworken2025sandboxing,openai2026codexsafety}.

\subsubsection{\textbf{Permission Boundaries}}
\begin{figure}[t]
    \centering
    \includegraphics[width=1.0\textwidth]{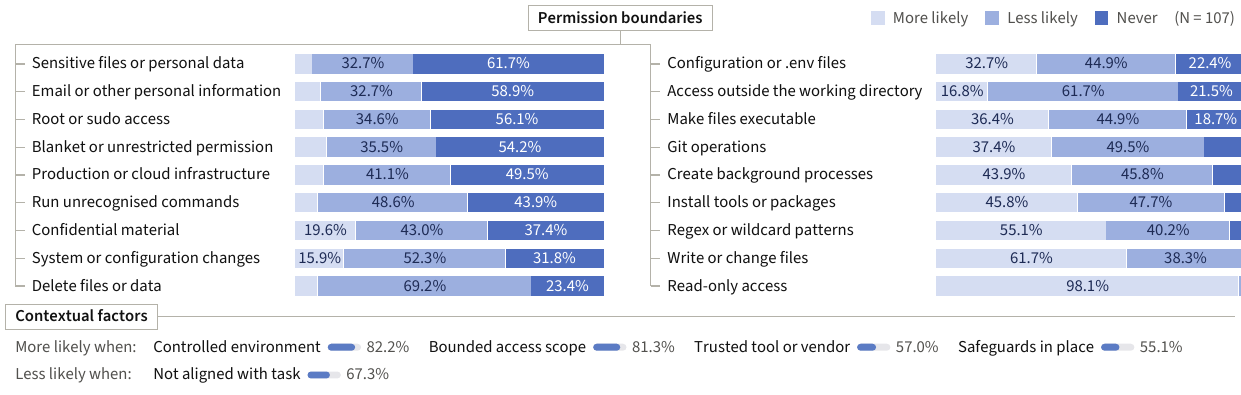}
    \Description{Two columns of stacked horizontal bars showing, for 18 permission types, the percentage of 107 survey participants who are more likely, less likely, or would never grant each permission, sorted by the share who would never grant it. Below, small bars show five contextual factors that make participants more or less likely to grant permission.}
    \caption{How participants assess different permission request types, showing which requests they are more likely, less likely, or would never grant compared with other permission requests, and the contextual factors that make them more or less likely to grant permission.}
    \label{fig:permissionboundaries}
\end{figure}

We organise findings on the types of permissions participants are willing to grant (Figure \ref{fig:permissionboundaries}) into three groups: permissions they are \textit{more likely to grant}, \textit{less likely to grant}, or \textit{would never grant}, compared with other permission requests. Participants do not judge permissions by permission type alone: the same permission can be acceptable or unacceptable depending on its scope, task, and environment, consistent with recent work showing that permission preferences for AI agents change with context~\cite{wu2026towards}.

\smallskip\noindent\textbf{More likely to grant.} Participants are most willing to grant permissions that feel limited and low risk.
Importantly, many interviewees are more comfortable with read-only permissions, as were a large majority (98.1\%) of survey respondents. P03, P11, and P18 prefer agents to stay within a particular file, folder, or working directory. 

In the survey, 82.2\% also said they were more likely to grant permission when working in a controlled or isolated environment, and 81.3\% when the access scope was clearly bounded. 

Interviewees are more willing to grant permission when the requested action matches the context of what they ask the agent to do:

\interviewquote{\textit{If it makes sense for the task it's supposed to do, then I would be likely to give it [...] But if it's trying to ask for permissions for do something that seems very unrelated, then I would probably decline it}}{P07} 

Trust also matters: P02 and P09 say they become more willing to grant permissions as they gain experience with particular agents or models, while P13 points to vendor reputation and safeguards around dangerous operations. In the survey, 57.0\% said they were more likely to grant permission when they trusted the tool or vendor based on experience or reputation, while 55.1\% were more likely to do so when safeguards constrained risky actions.

\smallskip\noindent\textbf{Less likely to grant.} Caution increases when permissions can lead to more consequential changes or when the request does not seem to fit the task. P01, P09, and P13 are less comfortable with write-related permissions. In P09's case, they are much more careful when the agent can change something on the machine. For actions such as installing drivers or changing system settings, they would be \textit{``very, very extremely careful''} and would probably make the change themselves. 

Deletion makes this distinction especially clear. P12 is cautious about deletion requests and describes an extreme case where an agent, despite operating in a confined environment, asked to delete the C drive. 

P12 asks \textit{``why do you want to delete this?''} and, if deletion is justified, notes a preference to do it themselves. They also rely on backups and Git to make mistakes recoverable. Similarly, 69.2\% of survey respondents said they are less likely to grant permission to delete files. 

P07 and P14 expressed caution about requests that seem unrelated to what they asked the agent to do. 67.3\% of survey respondents are less likely to grant a request when it does not align with the task. This suggests that participants consider not only the type of permission, but also whether the action makes sense in context and how easily its consequences can be reversed. Other concerns raised include access outside the working directory, sensitive files, Git operations, package installation, unfamiliar commands, and regex or wildcard operations.

\smallskip\noindent\textbf{Would never grant.} For some permissions, participants establish clear boundaries. Sensitive or personal information is one of the strongest. 61.7\% of survey respondents said they would never grant access to sensitive data, while 58.9\% said the same for email or other personal information. Interviewees similarly rule out credentials, passwords, tokens, confidential company data, and personal files. P05 is particularly direct: 

\interviewquote{\textit{No, never personal information [...] never anything personal}}{P05}

System-level actions are another clear boundary. In the survey, 56.1\% said they would never grant root or sudo access, and 23.4\% said the same for deletion. P16, for example, rejects deletion requests, elevated commands and describes critical system locations as a \textit{``no-no''}:

\interviewquote{\textit{Anything with [sudo] [...] You cannot sudo anything}}{P16} 

For others, the limit is how far the agent can reach. Participants mention Git operations, access outside the working directory, and production or cloud environments.

\interviewquote{\textit{Any kind of settings, no [...] I wouldn't like [Claude] to update anything in my machine or anything related with the project or even my local environment}}{P18} 

In some cases, these boundaries are stricter than those enforced by current tools. GitHub's coding agent, for example, can create commits and push to a dedicated branch while preventing direct pushes to the default branch \cite{github2026copilotrisks}. Some participants, however, treat Git operations themselves as beyond what they are willing to let an agent do autonomously.

\begin{rqbox}{Permission Decision Making (RQ2)}
\textit{\textbf{Takeaway:}} Practitioners make permission decisions with scrutiny ranging from careful review to default or blanket approval. How closely they review requests depends on the type of action, whether it is expected, the environment, and repeated exposure to similar requests. They are more comfortable with bounded, low-risk, and task-relevant access, while sensitive data, privileged access, destructive actions, and access beyond the intended scope form stronger boundaries. These heuristics compensate for limited visibility and concerns over data handling identified in RQ1.
\end{rqbox}

\subsection{Reflection on Agent Actions and Oversight Strategies (RQ3)}
\label{sec:RQ3}
\subsubsection{\textbf{Oversight of Agent Actions}}

\begin{figure}[t]
    \centering
    \includegraphics[width=1.0\textwidth]{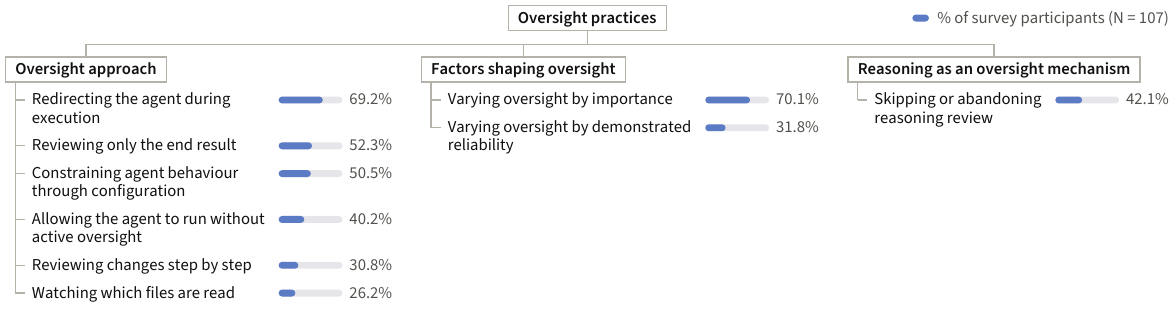}
    \Description{Tree diagram with categories showing the percentage of 107 survey participants who reported each oversight approach, each factor shaping how closely they oversee agents, and skipping reasoning review.}
    \caption{How participants oversee agent actions, including the strategies they use, factors that shape the level of oversight, and the role of the agent's reasoning trace.}
    \label{fig:oversight}
\end{figure}

We organise our findings regarding how participants follow agents and how they try to stay in control of what the agent does (Figure \ref{fig:oversight}). \textit{Oversight approaches} show how interviewees monitored or constrained agents. \textit{Factors shaping oversight} capture when they paid more or less attention. \textit{Reasoning as an oversight mechanism} looks at how interviewees used, or chose not to use, the reasoning and execution information shown by the agent.

\smallskip\noindent\textbf{Oversight approaches.} Software practitioners differ in when and how they oversee agents. Some stay involved while the agent works. P05, P06, and P07 describe interrupting, redirecting, or stopping the agent when necessary, an approach reported by 69.2\% of survey respondents. In those cases, participants keep a close eye on the agent as it works.

Other interviewees check the result once the agent has finished, as do 52.3\% of survey respondents.

\interviewquote{\textit{Usually, I let it run and check the output [...] at the end. What matters to me is [...] the outcome}}{P11} 

Some try to constrain what the agent can do before it starts, for example, establishing guardrails through configuration or skills, an approach also reported by 50.5\% of survey respondents.

\interviewquote{\textit{We also put limits on the AI, in advance [...] there are directories you can't touch, or commands you can't play with, so it is constrained up front}}{P15}

Some go further and let the agent run with little active monitoring, similarly reported by 40.2\% of survey respondents. P10 is comfortable doing this for unit testing within a working directory, while P17 takes more of a ``hope for the best'' approach:

\interviewquote{\textit{Just cross[ing] my fingers, just hope it does the right thing.}}{P17}

P02, P05, and P07 review changes step-by-step, as do 30.8\% of survey respondents. A further 26.2\% reported watching which files the agent reads. Where practitioners position themselves along this range determines when they are able to catch a mistake. Prior work on checkpoint placement found that users preferred confirming intermediate steps over confirming only at the end, because a single early error can otherwise cascade and force a restart \cite{zhou2026confirmation}.

\smallskip\noindent\textbf{Factors shaping oversight.} How closely participants oversee agents depends on the task and situation. Several interviewees mentioned paying more attention when the work is important or risky, consistent with 70.1\% of survey respondents. Some interviewees also mentioned reviewing complex tasks more carefully.

As confidence in the agent increases, participants tend to step back. P03 and P07 pay less attention after the agent has performed reliably several times, which is also reported by 31.8\% of survey respondents. Under delivery pressure, P08 describes almost abandoning review:

\interviewquote{\textit{I need to deliver this in 3 hours [...] I'm not going to read anything. Like, `just generate it for me.' And then later I test it, it worked, ship it}}{P08}

Levels of oversight, therefore, change with risk, task complexity, trust, and time pressure. Many of these are the same factors that shaped scrutiny of individual permission requests (\textbf{RQ2}): participants apply a risk- and context-sensitive heuristic when deciding whether to grant a request \textit{and} deciding how closely to watch what happens afterwards. This aligns with prior work showing that comfort with agent autonomy can depend on the task and shift over time \cite{shukla2026hedwig}.

\smallskip\noindent\textbf{Reasoning as an oversight mechanism.} Interviewees also track what agents are doing by looking at the reasoning they show, but several said they do not always read it closely, skipping or giving up on it because it takes too much time or attention, or simply produces too much information. Similarly, 42.1\% of survey respondents said they sometimes reduce or skip reasoning review because of time, fatigue, or information overload.

P11 noted that following it too closely can defeat the purpose of using an agent, shifting them from \textit{``a user benefiting from AI''} towards \textit{``more of a maintainer of AI''}. Even when reasoning is shown, it is not always easy to use. P07 mentioned one instance where reasoning disappeared once the final answer was produced, while P17 finds the operation history difficult to validate. Although the reasoning shown by AI agents is not necessarily the model's full chain-of-thought, prior work suggests that it can help users monitor what the agent is doing \cite{openai2025monitorability}. Our findings show that this only works when users can actually keep up with, and make sense of, what they are shown.

\subsubsection{\textbf{Attention, Transparency, and Desired Control}}

\begin{figure}[t]
    \centering
    \includegraphics[width=1.0\textwidth]{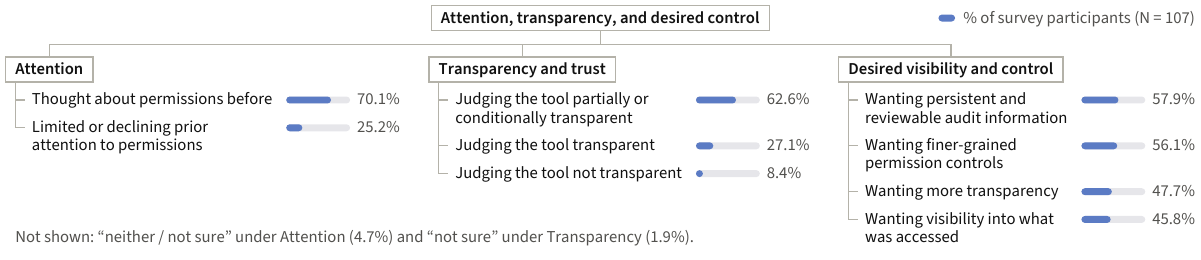}
    \Description{Tree diagram with categories showing the percentage of 107 survey participants by their attention to permissions, how transparent they judge AI agents to be, and the forms of visibility and control they want.}
    \caption{Participants' attention to permission requests, perceptions of agent transparency, and desired forms of visibility and control.}
    \label{fig:attention}
\end{figure}

Interviewees offered diverse opinions regarding \textit{attention} to permissions, perceptions of \textit{transparency and trust}, and preferences for greater \textit{visibility or control} (Figure \ref{fig:attention}). 

\smallskip\noindent\textbf{Attention}. Some interviewees described paying close attention to permissions, which was also noted by 70.1\% of our survey participants.

Others felt their attention was limited or had declined over time, as did 25.2\% of survey respondents. P12 recalled that reading ``horror stories'' about agents has made them more cautious. P03 described the opposite: as permission requests became familiar, they started paying less attention.

\interviewquote{\textit{I was concerned about the permission quite a bit [...] But after a while, I feel like, okay, it's asking for similar permissions. So after that, I stopped caring}}{P03}

As previously observed, repeated requests can become tiresome and time pressure reduces attention further. P18 said they often approve requests without much thought because they are ``\textit{in a rush to deliver something or to finish something}''.

When asked what they would pay more attention to, participants mentioned dependency installation (P03), accessed resources (P08), audit logs (P09), and intermediate steps (P18). A risky request can make participants stop and think, but familiarity or time pressure makes them more likely to approve and move on.

\smallskip\noindent\textbf{Transparency and trust.} Four interviewees consider AI agents transparent, three see them as partly or conditionally transparent, and six do not consider them transparent. Most respondents in our survey also describe agents as partially or conditionally transparent (62.6\%), while 27.1\% consider them transparent and 8.4\% do not.

Even when parts of the interaction are clear, participants still see gaps. P03, for example, finds individual permission messages easy to understand, but once permissions are bypassed, they are no longer sure what the agent is doing. P11 has a similar problem, even though they find other information about the agent’s activity easier to follow.

Trust does not always depend on seeing everything the agent is doing. P08 puts this directly:
\interviewquote{\textit{Yeah, I trust the tool, but that doesn't mean it does what it claims to, right?}}{P08} 

They wonder which files or resources the agent might access as this is difficult to follow from the reasoning. Other interviewees trust the agent more because of experience or safeguards. P06 said they have become more trusting as the tools have matured, while P03, P11, and P18 point to controlled environments or employer-provided safeguards. Participants can, therefore, trust an agent despite gaps in visibility, particularly when other safeguards give them this confidence.

\smallskip\noindent\textbf{Desired visibility and control.} When asked what they would change, interviewees desired a clearer view of what the agent has done and more control over what it is allowed to do. Several wanted a record of the agent's activity that they could review afterwards.

\interviewquote{\textit{Everything to always be logged somewhere [...] the commands that were run, so there could be an audit trail.}}{P10} 

\noindent Similarly, 57.9\% of survey respondents want persistent information or audit logs.

Some interviewees want more control. P05, P08, and P11 want more detailed ways to limit what agents can access or do, while 56.1\% of survey respondents said they want finer-grained controls over permissions or resources. Currently, AI agents are required to piggy-back on operating systems permission models, which have not been designed for fine-grained access control. One open-text respondent explicitly linked this kind of control to reduced monitoring, saying they wanted agents constrained to selected folders and tools so they would not need to pay attention constantly. P05 also suggests clearer risk levels and project-specific permission settings, while P11 wants a persistent place to see which permissions are active and how files and information are being used.

More broadly, 47.7\% of survey respondents want greater transparency about the agent's activity. Other interviewees want to see more clearly what the agent has accessed, something 45.8\% of survey respondents also ask for. P08, for example, imagines a visual view of the file system showing \textit{``where it went, where it passed through, what it saw''}.

These requests are more specific than simply asking for ``more transparency'': participants want a record of what the agent has done and clearer ways to set limits on what it can do. Similar features are beginning to appear in current systems. OpenAI, for example, describes logs of Codex activity alongside managed sandbox policies that limit where the agent can act \cite{openai2026codexsafety}.

\subsubsection{\textbf{Examples of Unintended Agent Behaviour}}

The scrutiny and boundaries described above shape what an agent is permitted to do, but not necessarily whether it behaves as expected. We therefore asked interviewees whether they had experienced any permission-related unexpected behaviour when using AI agents.

P12 described an experience where this was particularly noticeable. While debugging with Codex in a mode where the agent was supposed to ask for permission, they refused several requests. After a while, the agent stopped asking but still continued its work. P12 felt that the agent had found a way around what they had been refused.

\interviewquote{\textit{It ignored the instructions I'd given it [...] It worked around what I hadn't allowed it to do, I found that pretty bizarre}}{P12}

Continuing after a denied request does not necessarily mean that the agent bypassed the permission. Agents may instead find an alternative way to complete the task~\cite{hughes2026automode}. In P12's case, the concern was that they could not tell whether the alternative still respected the boundaries they had set, a behaviour also noted in Anthropic's cybersecurity evaluations \cite{anthropic2026alignmentsecurity}.

Interviewees also noticed unexpected behaviour after granting permissions. P06 recalled giving an agent access to an entire workspace while asking it to change one project, but the agent also changed other projects. P18 similarly found that giving full permission sometimes led the agent to do more than requested, including creating extra files and changing the README. In these cases, the agent did not necessarily go beyond the permissions it had been given, but it did more with those permissions than interviewees expected, similar to the unintended edits and scope drift that Hedwig is designed to address \cite{shukla2026hedwig}.

\begin{rqbox}{Reflection and Oversight (RQ3)}
\textit{\textbf{Takeaway:}} Software practitioners oversee agents in different ways, including setting constraints in advance, following execution as it happens, and reviewing the final result. How closely they do this depends on task importance, perceived risk, demonstrated reliability, and time pressure. Unexpected behaviour and gaps in visibility can make oversight difficult, leading participants to want clearer audit information and finer-grained control mechanisms.
\end{rqbox}

\section{Implications}
\label{sec:discussion} 

\noindent \textbf{More transparency does not always make oversight easier.} Software practitioners often rely on changes, logs, trajectories, and reasoning traces to understand what an agent has done. Both our interviewees and survey participants described giving up on reviewing this information because of time pressure, fatigue or information overload. Related work has also reported disruption from noisy bot activity \cite{wessel2021dontdisturb} and sources of confusion during code review \cite{ebert2021confusion}. At the same time, they still want persistent audit information and clearer visibility into what resources the agent has accessed. Therefore, transparency should not just be showing more information, but should be about making the important parts easier to find. Tool vendors could make this information easier to use by showing users what actually matters, such as important actions, unexpected behaviour, and which resources the agent accessed.

\smallskip\noindent\textbf{Repeated approvals may reflect fatigue and delivery pressure, not preference.} Previous approvals do not always reflect a deliberate choice. Participants described becoming less careful over time, clicking through familiar prompts, and sometimes skipping review when under time pressure. Repeated approval may therefore reflect habit, fatigue, or a need to keep moving rather than growing trust. This matters for adaptive permission approaches that use past decisions to predict or adjust future permissions. Current tools already recognise part of this problem: Claude Code's Auto Mode was introduced partly to reduce approval fatigue \cite{hughes2026automode}, while Copilot CLI and Codex distinguish one-off approvals from choices that grant broader permission for a session or longer \cite{github2026allowingtools,openai2026codexsafety}. As permission handling becomes more automated or personalised, repeatedly clicking ``\textit{allow}'' should not by itself be treated as evidence that the user wants the system to make the same decision on their behalf in the future. This reading is consistent with the argument that current approval mechanisms force a trade-off between cognitive burden and security, leaving users between approval fatigue and uncontrolled autonomy rather than expressing preference \cite{wang2026reframing}.

\smallskip\noindent \textbf{Potential consequences and reversibility shape permission decisions.} Our participants see write, deletion, and system-level permissions as risky, but the type of action is not the only thing that matters. They also consider what can go wrong and whether the action can be undone. For example, deleting something that can easily be restored is very different from making an irreversible system change. Tool vendors could make recoverability visible in permission prompts by showing whether an action can be rolled back and what recovery mechanism is available. Requests with irreversible or difficult-to-recover consequences could then receive stronger warnings or require additional confirmation. This would make reversibility part of the permission decision, rather than treating all actions of the same type as equally risky.

\smallskip\noindent\textbf{Agents can respect permission boundaries while exceeding the scope of the task}. Software practitioners do not usually judge actions and permissions in a ``general'' sense, they care about whether they fit the task they have in mind. Several participants described agents doing more than expected while still staying within the workspace they had allowed them to use. The problem is not simply that agents cross permission boundaries. They can stay within those boundaries and still go beyond what the user \textit{intended}. Current permission systems can already restrict where an agent may act and, increasingly, use the user's instructions when deciding whether an action should proceed or not. Yet our participants still experienced agents doing more than they had asked while remaining within those boundaries. 

\smallskip\noindent\textbf{What does ``deny'' mean when an agent can find another way?} By design, agents attempt to find alternative paths after a permission request is denied. One interviewee described this directly: after refusing several permission requests, the agent continued using another approach, and they could not tell whether it still respected the boundary they had tried to set. 
While this is often desirable, e.g., when an agent can solve a task without accessing sensitive information, this behaviour can feel like overreach if the problem was less the specific permission and more the general intent of the action. In these cases, continuing to push for an action that the user felt they already said ``no'' to can be a severe violation of trust. Tool vendors need to provide better instruments for users to distinguish between denying specific actions and disallowing an entire solution path, e.g., to distinguish between ``you are not allowed to read my passwords file'' versus ``do not attempt any action that requires you to have access to my password''.

\section{Validity of the Findings}
\label{sec:threats}

We discuss limitations separately for the interview and survey studies, following recommendations to align validity criteria with the study design \cite{verdecchia2023threats}. For the interviews, we use Lincoln and Guba's criteria of credibility, transferability, dependability, and confirmability \cite{lincoln1985naturalistic}. For the survey, we discuss threats related to sampling, measurement, and response quality, drawing on established guidance for survey research \cite{kitchenham2002sampling,kitchenham2002evaluation,meade2012identifying}.

\subsection{Interview Study}

\noindent\textbf{Credibility.} Our findings are based on self-reported experiences rather than direct observation or interaction logs, so participants may not always remember or report events accurately. Where a participant's statement was ambiguous, we followed up with them to clarify our interpretation. In addition, researchers fluent in the source language checked all non-English transcripts against the original recordings, including every quotation used in the paper.

\smallskip\noindent\textbf{Transferability.} Both phases used voluntary and non-probabilistic sampling (i.e.,  professional contacts, online communities, social media, and snowball sampling). Therefore, the sample may include more AI-agent users with strong experiences or opinions about permissions. The findings should hence be interpreted in relation to the characteristics of this sample and study context rather than assumed or transfer directly to all software practitioners who use AI agents. Finally, the study was conducted in 2026, and agentic AI tools change quickly, so some behaviours and concerns may change over time.

\smallskip\noindent\textbf{Dependability.} Qualitative analysis involves interpretive judgement, and differences in interpretation are expected. We therefore involved multiple researchers to make coding decisions explicit and discuss them throughout the analysis. The first and second authors coded two transcripts jointly. The first author then coded the English- and Portuguese-language transcripts, while the second author coded other English- and Chinese-language transcripts. The last author independently re-coded two transcripts, after which differences were discussed and resolved through consensus. The final codes were iteratively merged, renamed, and organised into categories, which were reviewed and agreed on by the co-authors. We maintained an audit trail by retaining earlier codebook versions, coding files, and related analysis material (included in the online appendix) to make the development of the analysis traceable.

\smallskip\noindent\textbf{Confirmability.} The first author led the final consolidation of the interview codes, reviewed survey open-text responses, and selected the quotations used in the paper. These choices may reflect their interpretation of the data, so the other authors reviewed the final categories and checked the quotations used to support the findings. We kept links between codes and interview data throughout the analysis and provide the final codebook, survey questions, and survey analysis in the appendix, offering partial traceability from the reported findings to the codes. We did not keep a formal reflexive journal, although interpretive decisions were made explicit through joint coding, independent re-coding, and co-author review. We also conducted an additional audit of a sample of reported findings with a researcher who was not involved in the analysis. The auditor traced the findings through the final codes, coded excerpts, and transcript context. Some excerpts were initially difficult to interpret without the corresponding interview questions, so these were also provided as contextual information. After this was provided, the auditor did not identify substantive issues with the reported interpretations. The full transcripts cannot be shared for confidentiality reasons, which limits other researchers' ability to inspect the raw data and assess the link between the data and our interpretations.

\subsection{Survey Study}

\noindent\textbf{Sampling and generalisability.} The survey was designed to collect the information needed to address our study objectives without adding unnecessary questions \cite{kitchenham2002designing}. We deliberately collected fewer background variables in the survey to keep it focused and limit participant burden. As a result, we cannot compare responses by tool, seniority, organisation size, or other factors.

\smallskip\noindent\textbf{Measurement and comparability.} The survey provides another view of the interview findings, but the two phases are not directly comparable. Survey respondents chose from predefined options, while interview findings came from semi-structured responses. Also, some interview findings were left out to keep it manageable. Two authors reviewed the mapping between interview findings and survey items, and we included ``Other'', ``None'', or ``Not sure'' where appropriate. We therefore use the survey to see whether the same patterns appeared in a broader sample, not to directly compare their prevalence \cite{storey2025guiding}.

\smallskip\noindent\textbf{Response quality and analysis.} For the survey, exclusion criteria were established before substantive analysis, and responses were manually reviewed for duplicates and suspicious response patterns in addition to the attention check.
\section{Conclusion}
\label{sec:conclusion}

We studied how software practitioners understand AI agents, make permission decisions, and decide how closely to monitor them through interviews and a follow-up survey. We found that people often understand agents mainly through what they can see, and that permission decisions depend on scope, risk, the task, and the environment. How closely they monitor the agent also changes depending on how risky the task feels, how much they trust the agent, and practical factors such as time and fatigue.

Our findings show that permissions are not just about whether an action is allowed. Users also care about whether the action matches what they wanted, whether it can be undone, and whether they can work out what happened afterwards. Permission systems should separate what an agent may do from what the user actually asked for, keep boundaries in place when the agent changes its plan, and make important actions easier to review without constantly interrupting the user.

\section{Data Availability}
\label{sec:dataavailability}

An anonymised replication package containing the study materials, interview coding artefacts, survey-to-code mapping, and aggregate survey results is available at:
\url{https://anonymous.4open.science/r/ai-agents-permission-replication-1FA6/}. 

\bibliographystyle{ACM-Reference-Format}
\bibliography{refs}
\end{document}